# Silicon nanoantennas for tailoring the optical properties of $MoS_2$ monolayers

Danae Katrisioti[1,2], Peter R. Wiecha[3], Aurélien Cuche[4], Sotiris Psilodimitrakopoulos[1], Guilhem Larrieu[3], Jonas Müller[3], Vincent Larrey[5], Bernhard Urbaszek[6], Xavier Marie[7,8], Emmanuel Stratakis[1], George Kioseoglou[1,2,*], Vincent Paillard[4], Jean-Marie Poumirol[4,*], and Ioannis Paradisanos[1,*]

*[1] Institute of Electronic Structure and Laser, Foundation for Research and Technology - Hellas, Heraklion, 71110, Crete, Greece*
*[2] Department of Materials Science and Engineering, University of Crete, Heraklion, 71003 Crete, Greece*
*[3] LAAS-CNRS, Université de Toulouse, 31000, Toulouse, France*
*[4] CEMES-CNRS, Université de Toulouse, Toulouse, France*
*[5] CEA-LETI, Université Grenoble-Alpes, Grenoble, France*
*[6] Institute of Condensed Matter Physics, Technische Universität Darmstadt, 64289, Darmstadt, Germany*
*[7] Université de Toulouse, INSA-CNRS-UPS, LPCNO, 135 Avenue Rangueil, 31077, Toulouse, France*
*[8] Institut Universitaire de France, 75231 Paris, France*

## *Abstract*

Silicon-based dielectric nanoantennas provide an effective platform for engineering light-matter interactions in van der Waals semiconductors. Here, we demonstrate near-field coupling between monolayer $MoS_2$ and silicon nanoantennas arranged in hexagonal lattices with tunable geometric parameters, leading to a 3-fold enhancement in photoluminescence and an excitation-wavelength-dependent emission that aligns with Mie-resonant modes. Raman spectroscopy reveals an up to 8-fold enhancement in the vibrational modes of $MoS_2$, while second-harmonic generation exhibits a 20 to 30-fold increase in efficiency, closely correlating with the presence of the underlying nanoantennas. Our experiments and simulations quantify the tunable benefits of the near-field interactions, taking into account thin-film interference and strain-induced effects. Our findings present dielectric nanoantennas as a promising platform for tailoring linear and nonlinear optical properties in 2D materials, with potential applications in nanophotonic devices and integrated photonics.

* Corresponding authors: gnk@materials.uoc.gr , jean-marie.poumirol@cemes.fr , iparad@iesl.forth.gr

Nanoscale light-matter interactions are key to advancing nanophotonics, enabling precise control over material optical properties[1]. Transition metal dichalcogenides (TMDs), atomically thin semiconductors, are ideal due to their strong excitonic effects[2,3], layer-dependent properties[4], mechanical strength[5] and crystal symmetry for applications in optoelectronics[6,7], quantum optics[8], nonlinear optics[9] and photonics[10]. Tailoring the optical response of TMDs using nanophotonic structures is a growing field with significant potential in technological applications. Approaches like plasmonic nanoparticles[11], metasurfaces[12], photonic crystals[13], and hybrid plasmonic-optical resonators[14] have been explored, however, they often come with inherent limitations such as high optical losses due to resistive heating in metals or low Q-factors[15]. In contrast, dielectric nanoantennas[16], made from high-refractive-index materials like silicon (Si)[17,18] or gallium phosphide (GaP)[19,20], support low-loss Mie resonances, enabling strong light confinement[21]. These antennas exhibit strong near-field interactions with adjacent TMDs, mediated by multipolar resonances that enable directional scattering, polarization control, and enhanced local field intensity[22,23,24,25,26,27]. Their compatibility with 2D materials like $MoS_2$, which can conform closely to nanoantenna surfaces[28], opens avenues for tuning light–TMD coupling across multiple optical processes.

In this work, we explore the near-field coupling between monolayer (E) $MoS_2$ and Si-based dielectric nanoantennas to modulate photoluminescence (PL), Raman scattering, and second-harmonic generation (SHG). We demonstrate a 3-fold PL enhancement with a 30 meV redshift (attributed to tensile strain), a Raman gain up to 8-fold depending on the excitation wavelength, and SHG enhancement up to 30-fold. These findings, supported by simulations, demonstrate how Si nanoantennas can be employed to engineer linear and nonlinear optical properties in 2D semiconductors, while also enabling us to disentangle the distinct physical origins of enhancement across different optical processes. Importantly, the present study goes beyond previous works that primarily focused on photoluminescence, by providing a systematic investigation of Raman scattering and second-harmonic generation in addition to PL. Through direct comparison with simulations, we offer a comprehensive picture of how near-field coupling in dielectric nanoantennas modulates both linear and nonlinear responses in monolayer $MoS_2$.

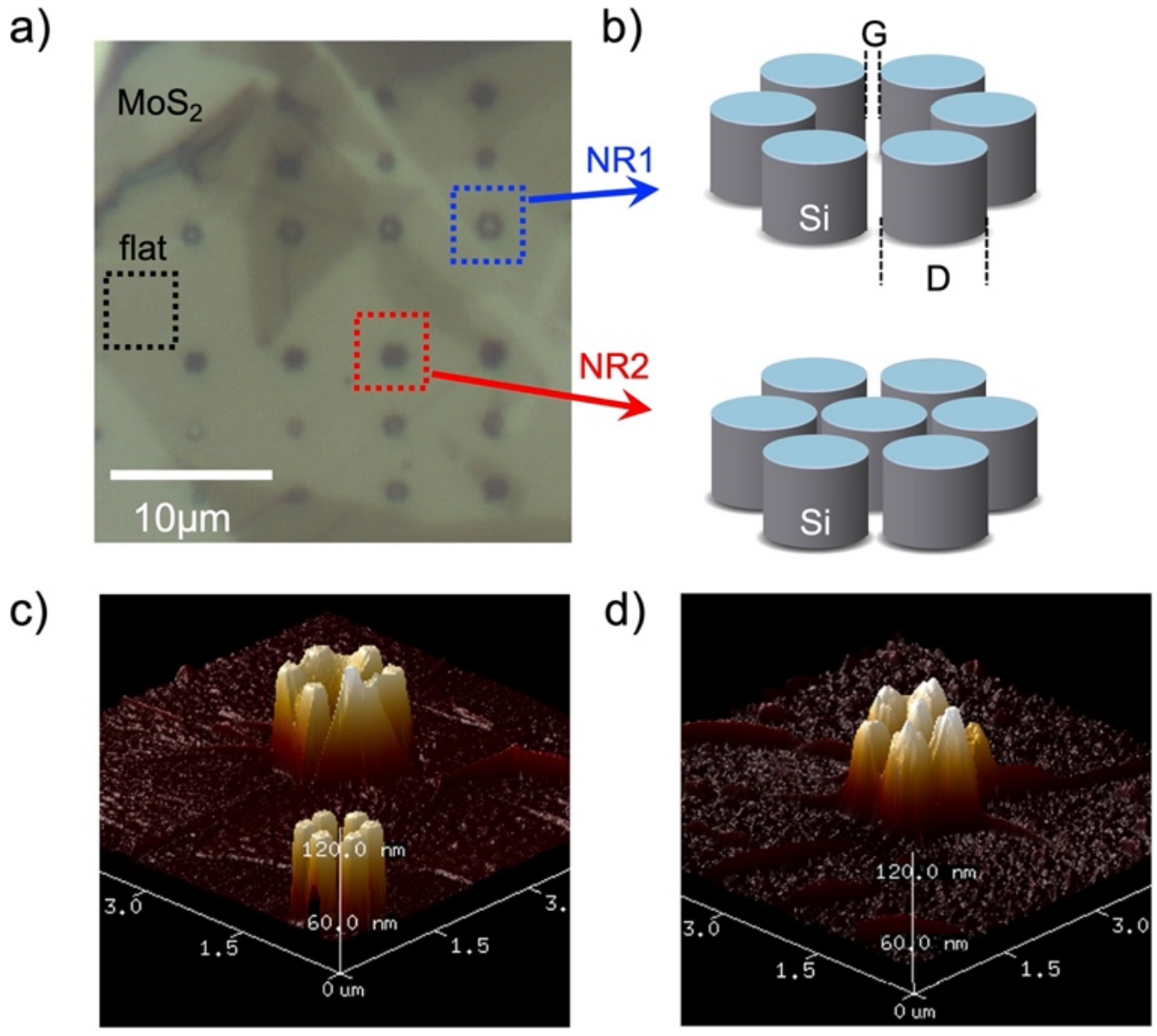

**Fig. 1.** (a) Optical microscope image of Si nanoantennas (NR1: blue; NR2: red) covered by 1L-$MoS_2$; flat $MoS_2$ region marked in black. (b) Schematic of NR1 (D = 300 nm) and NR2 (D = 250 nm, with central pillar); G = 300 nm for both. (c,d) 3D AFM images of NR1 and NR2, respectively, confirming full monolayer coverage; uncovered reference region visible below NR1.

Figure 1a shows a microscope image of two nanoantenna types -NR1 and NR2- on silicon-on-insulator (SOI) substrates. The arrays consist of cylindrical Si pillars fabricated via electron-beam lithography and anisotropic plasma etching (please, see Refs [29,30,31] for fabrication details), then capped with a ~30 nm $SiO_2$ layer. 1L-$MoS_2$ is mechanically exfoliated and transferred using a dry viscoelastic method[32]. NR1 is a hexameric array of Si pillars with diameter D=300 nm and gap G=300 nm. NR2 has a smaller diameter D=250 nm and includes a central pillar, forming a heptamer (Fig. 1a,b). Atomic force microscopy (AFM) images in Fig. 1c,d confirm full $MoS_2$ coverage on both types (see also supplementary material, Fig. S1). We focus on three regions:

flat $MoS_2$ (black square), NR1 (blue), and NR2 (red), as marked in Fig. 1a. The choice of a hexagonal arrangement was motivated by symmetry considerations, as it avoids anisotropy effects that could arise in rectangular or elliptical lattices, thereby minimizing polarization-dependent responses. The inter-pillar spacing (G = 300 nm) was selected to suppress strong pillar-to-pillar coupling, ensuring that the response originates primarily from the individual resonators. In addition, the footprint of each array is comparable to the diffraction-limited excitation spot in our spectroscopy experiments, which maximizes the collected signal. Comparing NR1 and NR2, which differ in pillar diameter and central filling, allows us to probe how subtle variations in geometry shift the near-field resonances, as confirmed by dark-field and PLE measurements (Fig. 2d).

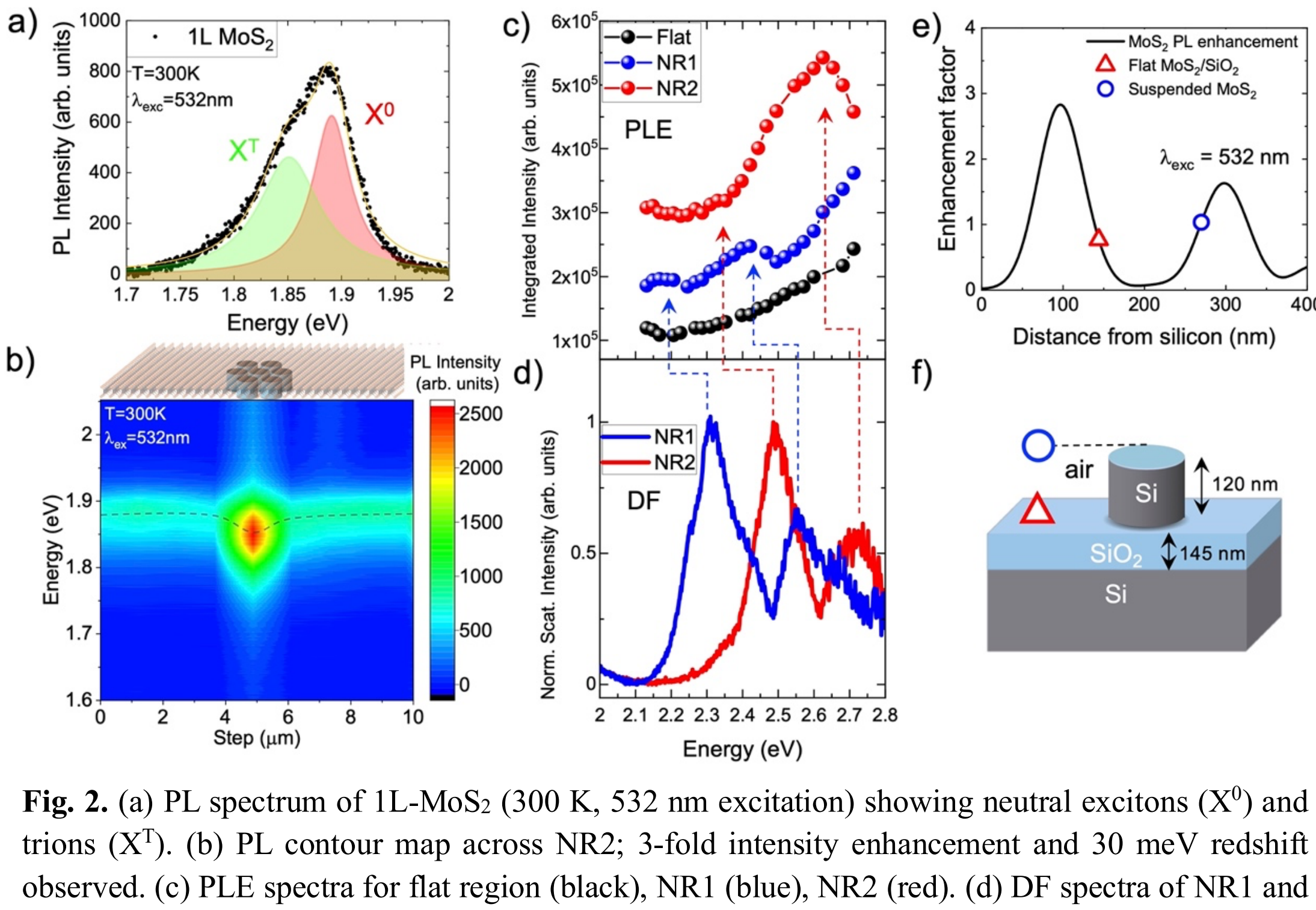

**Fig. 2.** (a) PL spectrum of 1L-$MoS_2$ (300 K, 532 nm excitation) showing neutral excitons ($X^0$) and trions ($X^T$). (b) PL contour map across NR2; 3-fold intensity enhancement and 30 meV redshift observed. (c) PLE spectra for flat region (black), NR1 (blue), NR2 (red). (d) DF spectra of NR1 and NR2, where scattered light was collected through a pinhole in the detection path and directed to a spectrometer instead of the imaging camera. (e) Simulated enhancement factor, $F_{total}$, based on multi-reflection model at 532 nm excitation and 660 nm emission. Different symbols correspond to different regions (flat and suspended), as indicated in (f) for the studied structure and represent simulation results rather than experimental averages.

We investigate the impact of Si nanoantennas on the excitonic response of 1L-$MoS_2$ using PL spectroscopy at 300 K. A representative PL spectrum from the flat region (Fig. 2a) shows distinct peaks for neutral excitons ($X^0$) and negatively charged trions ($X^T$), consistent with previous reports on 1L-$MoS_2$ on $SiO_2$ substrates[33,34] (for a wider spectral window, see supplementary material Fig. S2). Line scans across the nanoantennas (Fig. 2b) reveal a 3-fold PL intensity enhancement when the laser is centered on NR2 along with a 30 meV redshift. We directly quantify the tensile strain induced in 1L-$MoS_2$ by the nanopillars through the measured excitonic redshift, which corresponds to ~0.3% biaxial tensile strain. However, while strain is indeed present, its effect on the optical properties of $MoS_2$ is well understood: even small tensile strain lifts the Γ point above the K point, leading to momentum-indirect transitions and a reduction in PL efficiency[35,36]. Thus, strain alone cannot account for the observed ~3-fold PL enhancement on the nanoantennas. Instead, we attribute the enhanced emission primarily to near-field interactions, which increase the radiative recombination rate. Both NR1 and NR2 induce comparable levels of strain, but this strain contribution is secondary compared to the dominant near-field effects.

To explore the excitation-dependent absorption, we perform photoluminescence excitation (PLE) spectroscopy across the 460–585 nm (2.70–2.11 eV) range. To ensure non-resonant excitation conditions, we avoid tuning the laser near the B-exciton emission of 1L-$MoS_2$ at 600 nm (2.05 eV)[37]. While flat 1L-$MoS_2$ shows minimal spectral variation, NR1 and NR2 exhibit distinct PLE resonances (Fig. 2c), indicating coupling to Mie-type modes. Comparison with dark-field (DF) spectra (Fig. 2d) confirms that the observed PLE peaks arise from near-field effects. The redshift (~80–100 meV) relative to DF spectra reflects a common near-to-far-field spectral offset observed in nanostructures[30,38,39,40]. Given the height difference between the Si nanopillars and the surrounding $SiO_2$ flat regions (Fig. 1c,d), it is essential to assess the potential contribution of substrate-induced interference effects to the observed PL enhancement. To quantify these effects, we use a multi-reflection model based on Fresnel equations, following established methodologies and assuming a planar structure[41,42]. We calculate the total signal enhancement factor ($F_{\text{total}}$) as a function of the distance between 1L-$MoS_2$ and the underlying Si substrate (Fig. 2e,f), considering the complex refractive indices of 1L-$MoS_2$, air, $SiO_2$, and Si, the Fresnel transmittance and reflection coefficients, and the excitation and emission wavelengths. The enhancement factor is given by [41,42]:

$$F_{total} = N \int_0^{d1} |F_{excitation} F_{emission}|^2 dx$$

where $F_{\text{excitation}}$ and $F_{\text{emission}}$, correspond to the enhancement factors for the excitation light and emitted signal, respectively, integrated over the monolayer thickness (d1 = 0.63 nm). $N$ is a normalization factor, defined as the inverse of $F_{\text{total}}$ for a free-standing 1L-$MoS_2$, obtained by replacing the $SiO_2$ and Si layers with air. In Fig. 2e, we present the $F_{\text{total}}$ of the PL of 1L-$MoS_2$ as a function of the distance from the Si substrate under 532 nm (2.33 eV) excitation and 660 nm (1.88 eV) emission, allowing direct comparison with Fig. 2b. To evaluate the contribution of thin-film interference effects, we analyze two distinct regions using the multi-reflection model: flat 1L-$MoS_2$ (red triangle), and 1L-$MoS_2$ suspended between the Si-pillars (blue circles), as illustrated in Fig. 2f. From this comparison, we conclude that thin-film interference does not significantly contribute to the observed ~3-fold PL enhancement, as these two regions exhibit similar enhancement factors. This supports that the PL enhancement primarily originates from near-field coupling between 1L-$MoS_2$ excitons and resonant modes of the nanoantennas.

We now shift our focus to near-field enhancement mechanisms that do not involve exciton thermalization, radiative recombination, or non-radiative relaxation processes, but rather influence inelastic light scattering phenomena such as Raman scattering. We collect Raman spectra at selected excitation wavelengths: 473 nm (2.62 eV), 543 nm (2.28 eV) and 594 nm (2.09 eV). The E′ (in-plane) and $A_1'$ (out-of-plane) vibrational modes of 1L-$MoS_2$ are observed at 386 $cm^{-1}$ and 404 $cm^{-1}$, respectively, in agreement with previous studies [43,44]. Fig. 3a and 3b compare representative Raman spectra of 1L-$MoS_2$ on NR1 (blue) and NR2 (red) to that of flat 1L-$MoS_2$ (black). The Raman intensity exhibits significant enhancement when the laser spot excites 1L-$MoS_2$ on top of NR1 and NR2 at 543 nm (2.28 eV) and 473 nm (2.62 eV), respectively. Fig. 3c,d plots the integrated intensity ratio of the E′ and $A_1'$ vibrational modes, normalized by the intensity of flat 1L-$MoS_2$, for the three different excitation wavelengths. The results reveal that Raman scattering efficiency is enhanced by a factor between 2 and 8, depending on the excitation wavelength, with the maximum enhancement occurring in the green spectral region (543 nm, 2.28 eV). While NR1 and NR2 exhibit similar intensity ratios in the green and yellow spectral regions, NR2 demonstrates a slightly stronger enhancement in the blue region (473 nm, 2.62 eV). To interpret these findings, we calculate the spectra of the normalized local electric field enhancement for NR1 and NR2 single Si-pillars. We compare integrated Raman intensity (from experimental spectra) with field enhancement spectra using the Green's Dyadic Method (GDM) through our custom Python implementation, "pyGDM". The GDM is a volume integral method in the frequency domain. A nano-structure is discretized on a regular, hexagonal compact 3D grid, the coupling between all elements is described using Green's tensors. The presence of a substrate can be analytically included in the Green's tensors. More detailed information about the method and about our open source implementation can be found in Ref [45]. The pillars consist of a 95 nm silicon base (using tabulated refractive indices from literature [46]) and a 30 nm $SiO_2$ capping layer. We model the layered substrate using Green's tensors, comprising a bulk silicon base, a 145 nm $SiO_2$ spacer layer, and the Si nanopillar placed on top. The system is illuminated with a plane wave at normal incidence, spanning the same wavelength range as the experiments, and we incoherently average two orthogonal linear polarizations. The 300 nm inter-pillar gap in NR1 and NR2 allows us to treat the system as an incoherent sum of individual nanoantennas, as we found no significant contribution from pillar-to-pillar coupling. We calculate the electric field intensity enhancement just above the $SiO_2$ capping on top of the silicon pillar and we plot the spectra of the average near-field enhancement as dashed lines in Fig. 3c,d. The calculated spectra qualitatively reproduce the same trend as the measured normalized Raman intensity ratios, supporting the role of enhanced local near-field effects in modifying the Raman scattering cross-section. Note that although the calculated average near-field enhancement differs in energy between NR1 and NR2, as expected due to the different diameters of the Si-NRs, uncertainties arise in directly comparing the calculated energies of the near-field enhancement with the experimental Raman enhancement due to the discretization process and fabrication uncertainties (mainly due to the conformation of 1L-$MoS_2$ on top of the nanoantennas, see supplementary material Fig. S1), which can lead to spectral shifts of 10-20 nm in either direction (blue or red). Despite these uncertainties, the simulations for NR1 in Fig. 3c show good agreement with the experimental data, and the overall trends for NR2 still provide qualitative insights into the near-field enhancement mechanisms. We further simulate the spatial distribution of the near-field intensity enhancement in the 1L-$MoS_2$ plane for NR1 and NR2, using the excitation energies of the Raman experiments. These simulations offer deeper insights into the field distribution across each Si-pillar, as shown in Fig. 3e and 3f, which present a side-view of the electric field intensity distribution around the Si-pillar along with a normalized electric field intensity enhancement scale bar. Excitation at 2.10 eV (590 nm) results in a weak near-field enhancement at the top surface of both NR1 and NR2, showing qualitative agreement with the experimental observations at this energy. In contrast, excitation at 2.29 eV (541 nm) produces a stronger local field enhancement on top of the disc for both Si-NRs, while excitation at 2.64 eV (470 nm) leads to a more pronounced near-field intensity enhancement, primarily at the center of the NR2 surface. This enhanced field distribution possibly explains the higher Raman scattering intensity observed for NR2 at this energy.

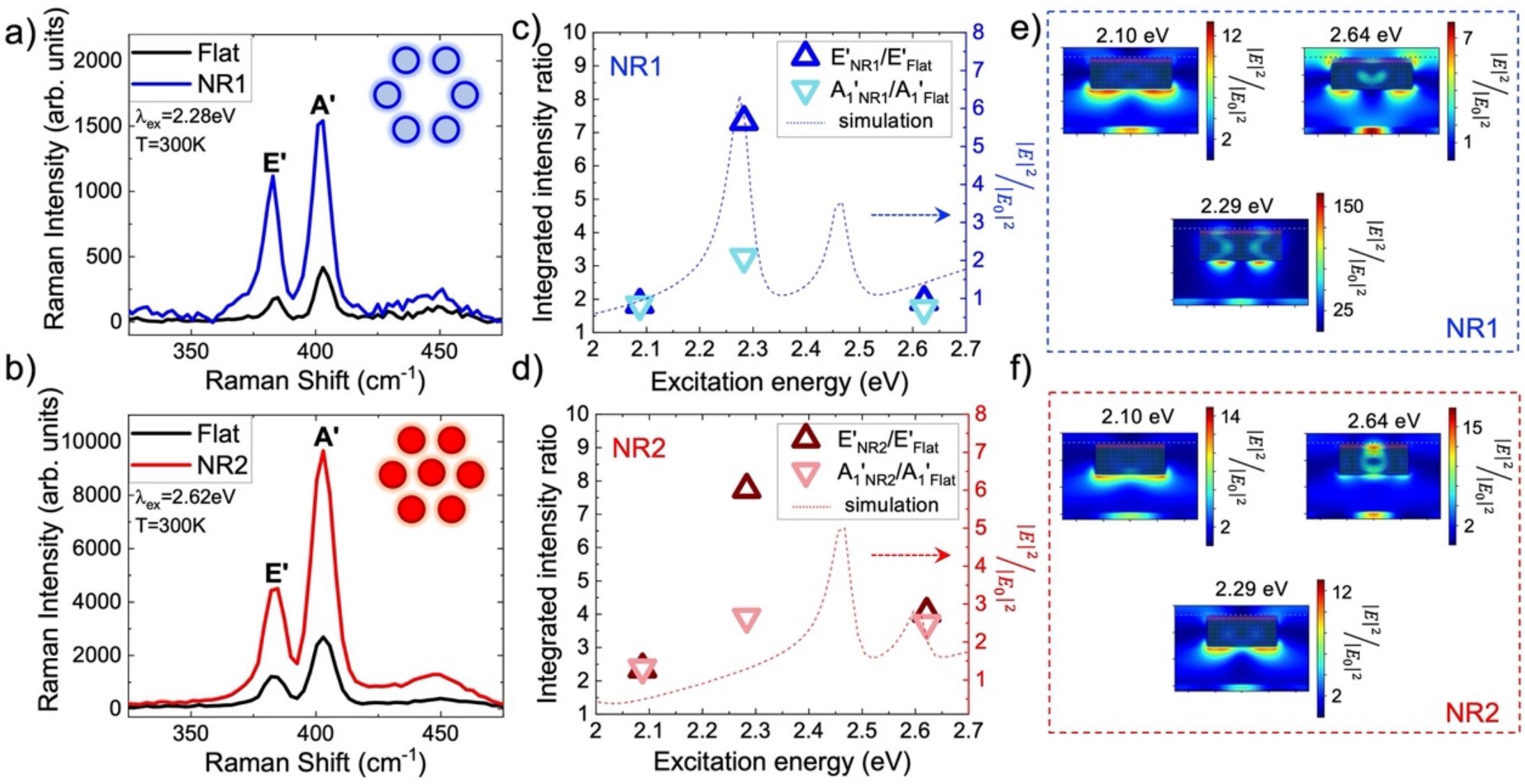

**Fig. 3.** (a) Raman spectra collected at T = 300 K using a 2.28 eV (543 nm) laser for NR1 (blue) and flat 1L-$MoS_2$ (black). (b) Raman spectra collected at T = 300 K using a 2.62 eV (473 nm) laser for NR2 (red) and flat 1L-$MoS_2$ (black). (c,d) Normalized Raman intensity of E' (dark blue) and $A_1'$ (light blue) for NR1 and NR2 at three excitation wavelengths; simulations shown as dashed lines. (e,f) Side-view simulations of near-field intensity distribution above NR1 and NR2 for excitation energies of 2.10 eV, 2.29 eV, and 2.64 eV.

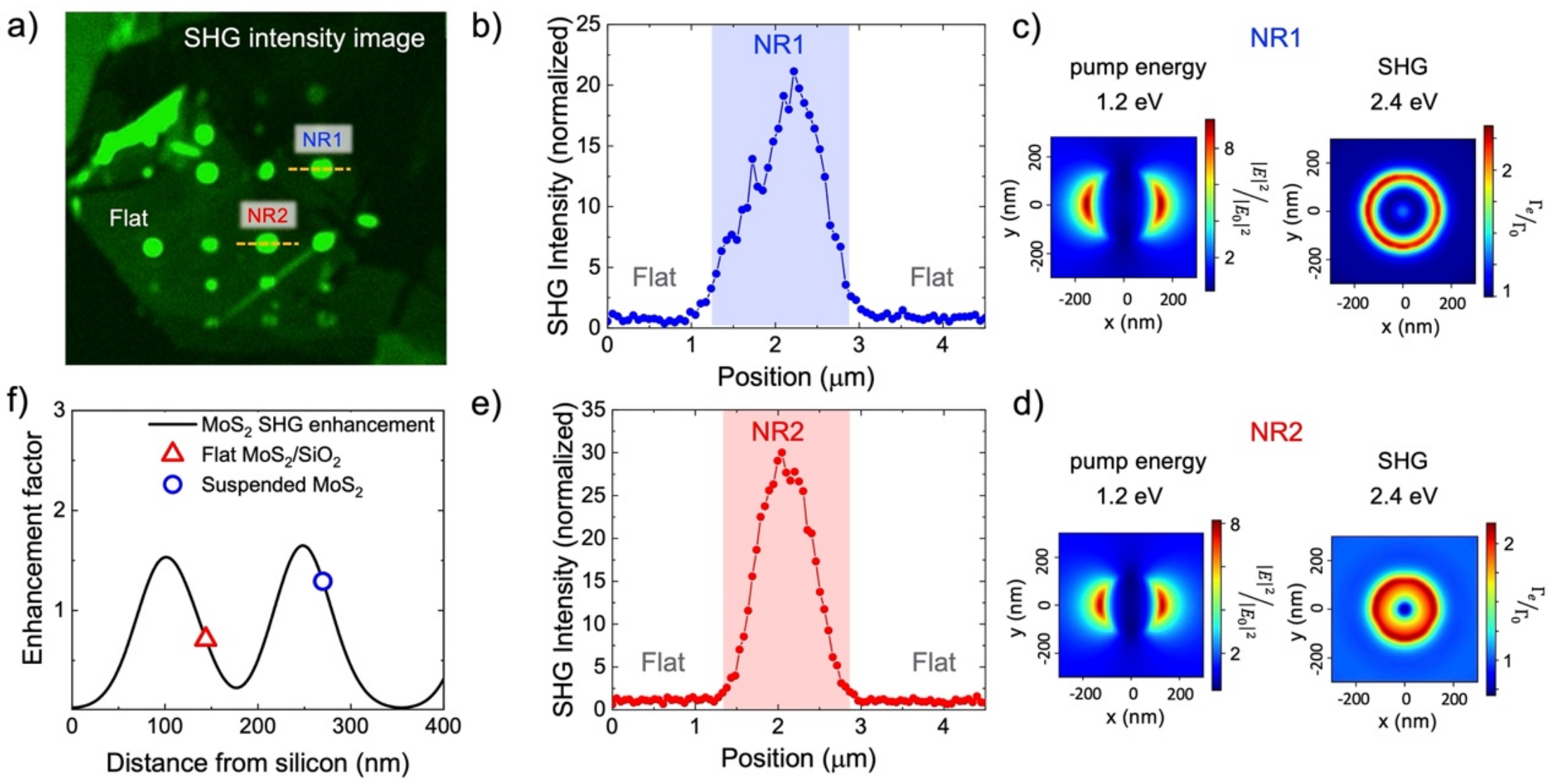

**Fig. 4.** (a) SHG intensity map ($\lambda_{pump}$ = 1030 nm), showing enhanced emission at NR1 and NR2 positions compared to flat 1L-$MoS_2$. (b,e) Cross-sectional SHG profiles (normalized to flat 1L-$MoS_2$) for NR1 and NR2, showing a 20-fold and 30-fold enhancement. (c,d) Simulated spatial distribution of the local electric field for a single pillar at a pump energy of 1.2 eV (left) and LDOS (right) for NR1 and NR2 at pump (1.2 eV) and SHG (2.4 eV) energies. (f) Classical interference model predicting SHG intensity for planar geometry. The pump energy at 1.2 eV is also considered in the model. The triangle and circle points represent simulation results rather than experimental averages.

Having explored the enhancement of PL and Raman scattering, we now turn our attention to the nonlinear optical response of 1L-$MoS_2$ coupled to Si-NRs, focusing on the enhancement of SHG. TMD monolayers belong to the $D_{3h}$ point group, which lacks inversion symmetry, and show intrinsically strong SHG signals[47,48]. However, a key challenge in utilizing SHG in ultrathin materials lies in the low absolute signal intensity, despite the exceptionally high nonlinear susceptibility of a single monolayer per unit thickness[49]. Recent studies demonstrate that resonant dielectric metasurfaces can be employed to engineer the amplitude and directionality of SHG through carefully designed mode coupling[50,51]. Our approach instead focuses on individual Si-NRs, where near-field effects drive localized SHG enhancement. We perform SHG imaging using a fundamental wavelength of 1030 nm (1.2 eV)[52] and the SHG signal is collected at 515 nm (2.4 eV), which is resonant or quasi-resonant with the majority of the Si-NRs fabricated in this study. Fig. 4a presents SHG intensity imaging normalized to the background, revealing uniform SHG intensity in flat 1L-$MoS_2$ regions but a clear enhancement at the precise locations of the Si-NRs, with variations in brightness across different nanoantennas. Examining the cross-sectional SHG intensity profiles for NR1 (blue) and NR2 (red), we measure a striking ~20 to 30-fold enhancement in SHG intensity compared to flat 1L-$MoS_2$ (Fig. 4b,e). Here, flat 1L-$MoS_2$ is normalized to 1.

The near-field intensity enhancement and the local density of optical states (LDOS) provide complementary insights into the nonlinear optical response of 1L-$MoS_2$/Si-NRs, with the former quantifying the local electric field enhancement at the pump energy (1.2 eV) and the latter indicating the availability of photonic states for emission at the SHG energy. In Fig. 4c and 4d, simulations using GDM reveal the top-view spatial distribution of these quantities for NR1 and NR2, respectively, in the plane of the TMD. At the pump energy of 1.2 eV, the near-field intensity enhancement exhibits a dipole-like pattern around the Si-NRs for both NR1 and NR2, indicating strong field localization that enhances the nonlinear polarization in the 1L-$MoS_2$ layer. Due to the second-order susceptibility ($\chi^{(2)}$) process governing SHG, the pump energy's contribution to the nonlinear response depends nonlinearly on the near-field intensity enhancement, with the SHG intensity scaling as the square of the local electric field. At the SHG energy of 2.4 eV, the LDOS in the TMD plane shows a distinct enhancement that facilitates SHG emission. These simulated field distributions and LDOS profiles qualitatively agree with the experimental findings in Fig. 3a, b, and e, where a 20- to 30-fold SHG enhancement is observed for 1L-$MoS_2$ on NR1 and NR2 compared to the flat region, as the enhanced near-field and increased LDOS together promote a more efficient nonlinear optical response. To evaluate contributions from classical thin-film interference effects, we apply again the multi-reflection model, modified for the specific conditions of SHG signal at 515 nm (2.4 eV), as well as the fundamental wavelength at 1030 nm (Fig. 4f). Suspended 1L-$MoS_2$ between the pillars is calculated to only exhibit a slightly larger enhancement compared to flat 1L-$MoS_2$ ruling out significant contributions from multi-reflections of propagating waves in the substrate. Additionally, we rule out strain as a contributing factor, as previous studies indicate that tensile strain in TMD monolayers has little to no impact on the total SHG intensity (measured without polarization selection) and, in some cases, may even lead to a reduction in the signal[53,54].

Finally, we comment that the different near-field enhancements observed in PL (~3-fold), Raman (2 to 8-fold), and SHG (~20 to 30-fold) stem from their distinct physical mechanisms and interactions with the near-field. PL in 1L-$MoS_2$ is governed by exciton thermalization, involving both radiative and non-radiative recombination. Although Mie resonances enhance absorption -leading to a higher density of photogenerated excitons- the PL increase remains limited, likely due to excitation density-dependent dynamics[55] and Auger effects[56]. That is, absorption may be enhanced by a factor of 8, but PL rises only by a factor of 3. Prior studies show that in 1L-$MoS_2$, exciton lifetimes at 300 K are mainly limited by Auger recombination when exciton densities exceed $10^9$ $cm^{-2}$[57], a condition met in our experiments. Raman scattering, driven by inelastic light-phonon interactions, shows stronger enhancement than PL. SHG, a parametric nonlinear process confined to a nanoscale interaction volume, exhibits second-order dependence on the local field at the pump energy and the LDOS at the SHG energy-factors that likely explain its greater enhancement.

Our study demonstrates near-field coupling effects that modify photoluminescence, Raman scattering, and second-harmonic generation processes in $MoS_2$ monolayers transferred on hexagonally arranged Si nanoantennas. PL spectroscopy reveals a 3-fold enhancement in the photoluminescence intensity of 1L-$MoS_2$ on Si-NRs, accompanied by a 30 meV redshift indicative of tensile strain. However, strain does not contribute to the observed enhancement. Instead, PLE results provide evidence that the enhancement arises from the coupling of 1L-$MoS_2$ to the optical near-field of the Si-NRs at specific excitation energies. Raman spectroscopy measurements show a 2 to 8-fold enhancement of the $E'$ and $A_1'$ vibrational modes with a maximum at 2.28 eV excitation. Numerical simulations of local field distributions confirm this enhancement, supporting the role of near-field interactions in modifying inelastic photon-phonon interactions. Similarly, SHG exhibits a significant ~20 to 30-fold enhancement at 2.4 eV for 1L-$MoS_2$ on the Si-NRs compared to the flat region, confirming the dominant role of near-field coupling in driving this enhancement, as supported by simulations that reveal strong near-field intensity enhancement at the pump energy and a more than 2-fold increase of the LDOS at the SHG energy. The varying enhancement factors, ~3-fold for PL, ~2 to 8 fold for Raman, and ~20 to 30-fold for SHG, highlight the different physical mechanisms and their coupling with nanoantennas. While PL is influenced by exciton thermalization and recombination, Raman scattering is dictated by inelastic photon-phonon interactions, and SHG, as a parametric nonlinear process confined to a small interaction volume, exhibits the strongest dependence on localized field enhancements. These findings establish silicon nanoantennas as a scalable and efficient platform for tailoring the optical properties of 2D materials via near-field interactions, with potential applications in integrated photonics, nonlinear optics, and quantum technologies.

ACKNOWLEDGEMENTS

I.P. and D.K. acknowledge financial support by the Hellenic Foundation for Research and Innovation (H.F.R.I.) under the "3rd Call for H.F.R.I. Research Projects to support Post-Doctoral Researchers" (Project Number: 7898). S.P. acknowledges financial support by the Hellenic Foundation for Research and Innovation (H.F.R.I.) under the action "Basic Research Financing (Horizontal support for all Sciences), National Recovery and Resilience Plan (Greece 2.0)" (Project Number: 014772 – Project Acronym: MAYA). E.S. and G.K. acknowledge support by the EU-funded DYNASTY Project, ID: 101079179, under the Horizon Europe framework programme.

Calculations were performed using the massively parallel computing center CALMIP in Toulouse (project P1107). P.R.W. acknowledges support by the French Agence Nationale de la Recherche (ANR) under grant ANR-22-CE24-0002 (project NAINOS). We are grateful to Ms. Katerina Tsagaraki from the micro/nanoelectronics group (MRG) of FO.R.T.H. for the AFM measurements.

# Supplementary Material

S1: AFM microscopy of $1L\text{-}MoS_2$ on top of hexamers and heptamers.

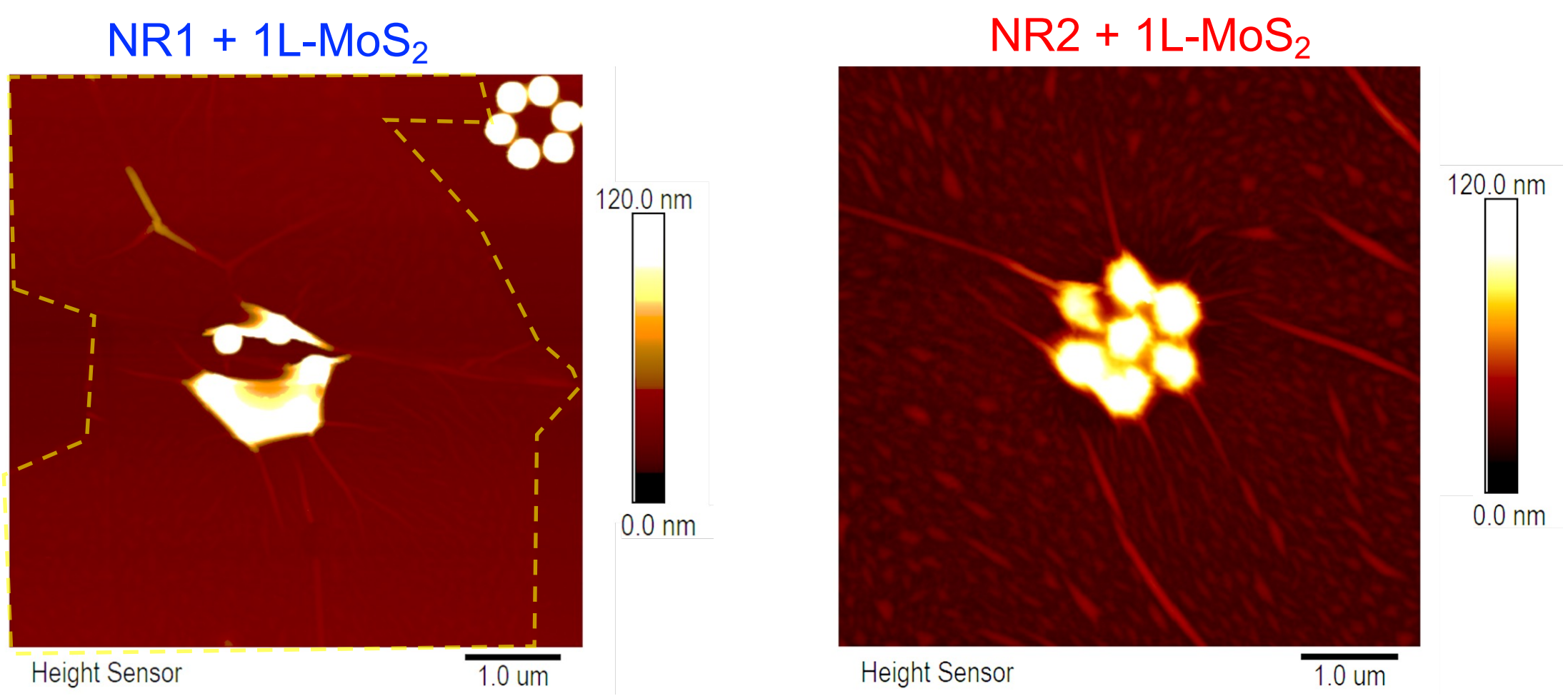

Fig. S1: 2D AFM images of $1L\text{-}MoS_2$ conforming to a hexamer (NR1, left) and a heptamer (NR2, right). For reference, a bare hexamer without $MoS_2$ coverage is shown in the top-right corner of the NR1 panel. The images also reveal the random formation of wrinkles in the flat monolayer and around the nanostructure.

S2: Photoluminescence spectroscopy of $1L\text{-}MoS_2$ excited by a 473nm laser.

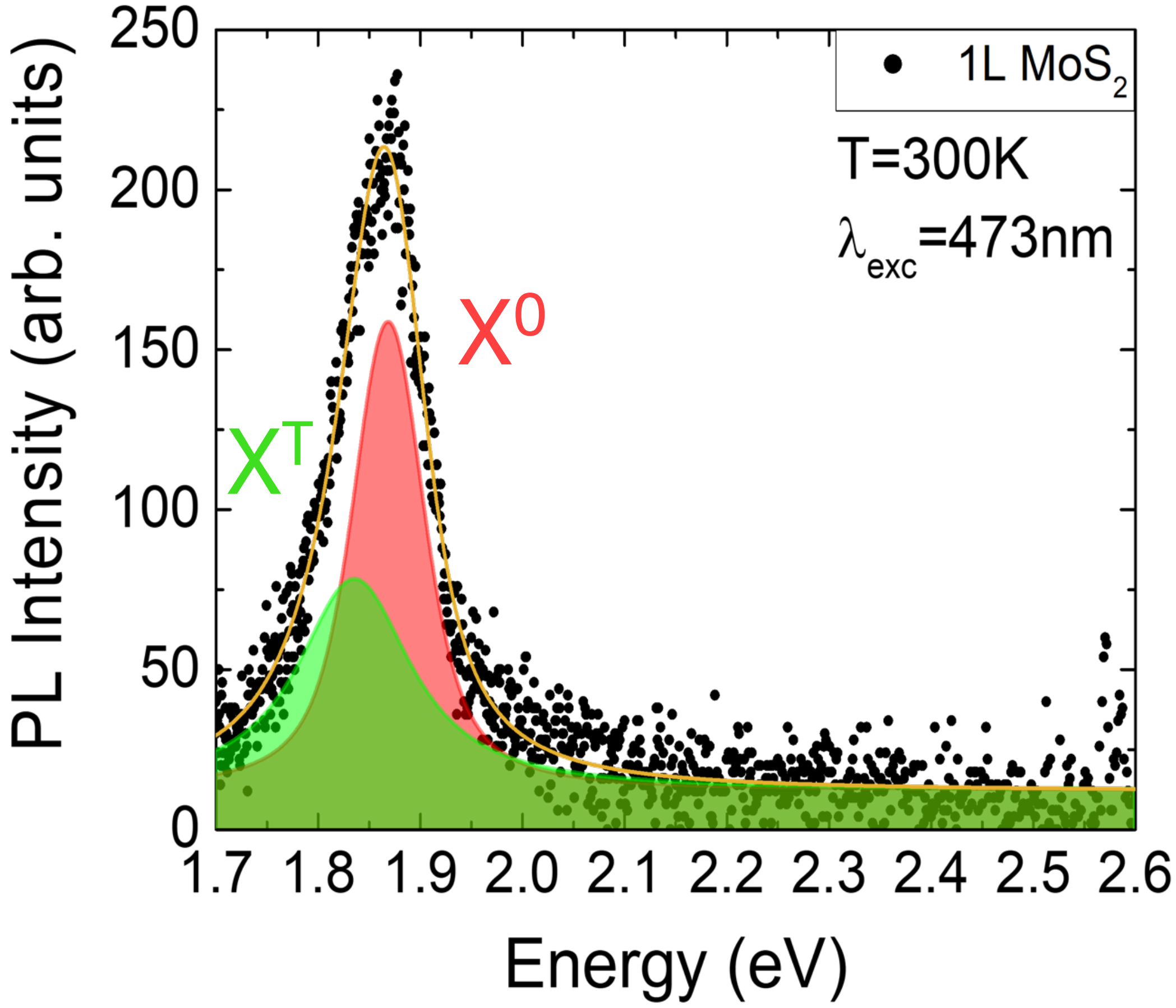

Fig. S2: Photoluminescence spectrum of 1L-$MoS_2$, under 473nm excitation (2.62 eV). Besides excitons ($X^0$) and trions ($X^T$), the spectrum does not show additional features at higher energies in the collection window.